\documentclass[twocolumn,showpacs,preprintnumbers,prd,superscriptaddress,nofootinbib]{revtex4}
\usepackage{amsmath}
\usepackage{amsfonts}
\usepackage{amssymb}
\usepackage{graphicx}
\usepackage[titletoc]{appendix}
\usepackage{color}

\graphicspath{{Graphics/}}

\newcommand{\tabref}[1]{Tab.~\ref{#1}}

\newcommand{\eqeqref}[1]{Eq.~\eqref{#1}}
\newcommand{\eqsref}[1]{Eqs.~\eqref{#1}}
\newcommand{\secref}[1]{Section~\ref{#1}}

\begin{document}

\title{Cosmographic analysis of the equation of state of the universe through Pad\'e approximations}


\author{Christine Gruber}
\email{chrisigruber@physik.fu-berlin.de}
\affiliation{Institut f\"ur Theoretische Physik, Freie Universit\"at Berlin, Arnimallee 14, D-14195 Berlin, Germany,}

\author{Orlando Luongo}
\email{luongo@na.infn.it}
\affiliation{Dipartimento di Fisica, Universit\`a di Napoli "Federico II", Via Cinthia, I-80126, Napoli, Italy,}
\affiliation{INFN Sez. di Napoli, Compl. Univ. Monte S. Angelo Ed. N, Via Cinthia, I-0126, Napoli, Italy,}
\affiliation{Instituto de Ciencias Nucleares, Universidad Nacional Aut\'onoma de M\'exico, AP 70543, M\'exico, DF 04510, Mexico.}

\date{\today}

\begin{abstract}
Cosmography is used in cosmological data processing in order to constrain the kinematics of the
universe in a model-independent way, providing an objective means to evaluate the agreement of a model with
observations. In this paper, we extend the conventional methodology of cosmography employing Taylor expansions
of observables by an alternative approach using Pad\'e approximations. Due to the superior convergence properties
of Pad\'e expansions, it is possible to improve the fitting analysis to obtain numerical values for the parameters
of the cosmographic series. From the results, we can derive the equation of state parameter of the universe and its
first derivative and thus acquire information about the thermodynamic state of the universe. We carry out statistical
analyses using observations of the distance modulus of type 1a supernovae, provided by the union 2.1 compilation of the
supernova cosmology project, employing a Markov chain Monte Carlo approach with an implemented Metropolis algorithm. We
compare the results of the original Taylor approach to the newly introduced Pad\'e formalism. The analyses show that
experimental data constrain the observable universe well, finding an accelerating
universe and a positive jerk parameter. We demonstrate that the Pad\'e convergence radii are greater than standard Taylor
convergence radii, and infer a lower limit on the acceleration of the universe solely by requiring the
positivity of the Pad\'e expansion. We obtain fairly good agreement with the Planck results, confirming the $\Lambda$CDM
model at small redshifts, although we cannot exclude a dark energy density varying in time with negligible speed of sound.
\end{abstract}

\maketitle

\section{Introduction}

Over the last decades, several experiments have gathered significant evidence suggesting
that our universe is currently undergoing an accelerated phase of expansion~\cite{uno,unobis}. Indications
are coming from the detection of type Ia supernovae (SNeIa)~\cite{SNeIa-1,SNeIa-4,Suzuki:2011hu}, from
measurements of the Hubble space telescope, from galaxy redshift surveys, cosmic microwave background
detection, baryonic acoustic oscillations and so forth (see~\cite{due} and references therein). A wide
number of theoretical models has been investigated in order to clarify the physical origin of such a
cosmic speed up \cite{g1000}, although so far no self-consistent solution has been found. Among the multitude
of approaches, ideas range from postulating a new ingredient dubbed dark energy (DE) driving the acceleration,
to modifications of the spacetime geometry of the universe itself~\cite{DE-Review}. A widely accepted and overall
quite successful framework is the so-called $\Lambda$CDM model~\cite{2006Cope}, until now the standard
paradigm of cosmology. Here, CDM stands for cold dark matter, and $\Lambda$ represents a cosmological constant~\cite{darkla}.
Unfortunately the model comprises two precarious issues, namely the problems of \emph{fine tuning} and
\emph{coincidence}, which do not allow us to consider $\Lambda$CDM as the final paradigm for describing the universe's
dynamics~\cite{1991}. The fine tuning problem refers to the large difference between the energy density driving
the cosmic expansion and the value for the vacuum energy density predicted by quantum field theory~\cite{1989Wein}.
The coincidence problem addresses the fact that the densities of the fluid driving the cosmic acceleration and
pressure-less matter are comparatively similar at the present time. In other words, although these two quantities evolved
differently during the history of the universe, their contributions to the overall energy density of the
universe are of the same order of magnitude today~\cite{wum,kunzo}. In the absence of a self-consistent theoretical
scheme to explain DE, the search for new cosmological models able to overcome both the fine tuning and coincidence
problems is an open task of modern cosmology. Any viable model attempting to describe the dynamics of the universe
must also provide agreement with the most recent observations~\cite{Planck}. As $\Lambda$CDM is already quite successful
in doing so, there is the common consensus that any feasible new model should reproduce the effects of
$\Lambda$CDM in the low redshift regime. Thus arises the conclusion that $\Lambda$CDM may be viewed as a first
approximation of a more complicated paradigm~\cite{capozziello,capozziello2,capozziello3}. Literature presents us with
a huge amount of models~\cite{MODELS,MODELS2} with diverse approaches, aiming at improving the shortcomings of
$\Lambda$CDM and explaining the aforementioned problems. Many of them are quite successful in describing the
observed phenomena, but with the growing number of models it becomes increasingly difficult to fairly discriminate
between them and favor a particular one over others. Indeed, the biggest problem of confronting models with data
is the fact that in such analyses the model in question is usually \emph{a priori} postulated to be the
correct one~\cite{correct}. This leads to a degeneracy among models, since the comparison of every model
to data favors the one being tested. Thus, many models seem to work better than others, leading to a desperate
need of independent methods to test cosmological paradigms~\cite{dege}. To this end, increasing efforts have been
devoted to the development of the so called cosmography of the universe~\cite{2008Wein,2005Viss}. Cosmography represents
the branch of cosmology attempting to obtain insights into the cosmological picture by exploring only the universe
kinematics, relying on as few assumptions as possible in order to keep a viewpoint as neutral as possible~\cite{2007Catt}.
We assume the validity of the cosmological principle only, i.e. the universe is supposed to be homogeneous and isotropic.
The dynamics of the universe can then be formulated in terms of a Friedmann-Robertson-Walker (FRW) metric,
$ds^2=dt^2-a(t)^2\left(\frac{dr^2}{1-kr^2}+r^2\sin^{2}\theta d\phi^2+r^2d\theta^2\right)$~\cite{2012Avil,25000}. The
methodology of cosmography is essentially based on expanding measurable cosmological quantities into Taylor series around
the present time, providing model-independent constraints on the universe's kinematics or energy density~\cite{cosmo}.

The present work's main purpose is to introduce an alternative technique to Taylor expansions to the framework of
cosmography. In particular, in order to carry out the cosmographic analyses, the formalism of Pad\'e approximations (PAs)
is proposed~\cite{padesx}. The use of PAs eliminates the convergence problem, e.g. the systematic errors due to truncated
Taylor series, which does not permit cosmography to provide reliable constraints at higher redshifts. Efforts to alleviate
the convergence problem have been employing the reparametrization of the redshift $z$ with \emph{ad hoc} auxiliary variables,
compressing data to shorter intervals of redshift where convergence of Taylor series is given~\cite{2012Avil}. Instead of using
artificial redshift constructions, we demonstrate that the PAs may be a viable alternative to improve the cosmographic fitting
procedure. Indeed, the convergence radii of PAs exceed those of the Taylor approach, justifying the use of data from a larger
redshift regime.

The paper is organized as follows: in~\secref{sec:cosmography} we give a general introduction of the
fundamental principles of cosmography and describe the technicalities of the analysis, as well as define
the conventional fitting functions in the Taylor formalism used in the literature. We shortly sketch how to
obtain certain important cosmological quantities that can be derived from the fitting parameters, like the
equation of state parameter of the universe. We will address some concerns about problems occurring in the
cosmographic procedure, and address the question of how to alleviate them in~\secref{sec:problems}.
In~\secref{sec:Pade}, introducing the concept of Pad\'e approximations as a way to circumvent the convergence
problem, we will describe in detail how to calculate the functions for the fits in the Pad\'e formalism,
and state their explicit form. Furthermore the convergence radius of a Pad\'e expansion will be introduced as
a quantitative measure of the range of validity of the approximation with respect to a Taylor expansion.
In~\secref{sec:results}, after devoting some attention to describing the statistical methods used, we will present
the numerical results of the analyses, i.e. the obtained values for the parameters of the CS, as well as the results for
the convergence radius of the Pad\'e expansion and the equation of state of the universe obtained from those values.
In~\secref{sec:concl} we will conclude our work.

\section{Cosmography of the universe}
\label{sec:cosmography}

The present section is devoted to describing in detail the role played by cosmography in the analysis
of the observable universe. As already mentioned before, the main feature of cosmography is its aim
to rely on as few underlying assumptions as possible. In particular, it is based on the validity of the
cosmological principle. In this work we will further use the assumption of a spatially flat universe,
i.e. $k=0$.

The standard procedure is to expand the scale factor $a(t)$ parameterizing the expansion of the universe
in the FRW-metric into a Taylor series with respect to time $t$ around the present time $t_0$, as
\begin{equation} \label{aint}
    a(t) \equiv 1+\sum_{\kappa=1}^{\infty}\frac{1}{\kappa !}\frac{d^\kappa a}{dt^\kappa}\Big|_{t=t_0}(t-t_0)^\kappa \,,
\end{equation}
with the coefficients in terms of scale factor derivatives evaluated at current time. \\
From these coefficients, we can define a set of parameters, given for a generic time $t$ as
\begin{subequations} \label{eq:CScoeff}
\begin{align}
    \mathcal H &= \frac{1}{a}\frac{da}{dt}\,, \\
    q &= -\frac{1}{a \mathcal H^2} \frac{d^2a}{dt^2}\,, \\
    j &= \frac{1}{a \mathcal H^3} \frac{d^3a}{dt^3}\,.
\end{align}
\end{subequations}
They are known in the literature as the Hubble parameter, the acceleration parameter and the
jerk parameter, respectively~\cite{riferimento}, and evaluated at the present time $t_0$
customarily termed the cosmographic series (CS)~\cite{2008Wein,2005Viss,2007Catt}. The CS
can be extended to higher orders defining further parameters, but in this work we restrict
ourselves to the first three. Each parameter of the CS has its distinct physical meaning. The
acceleration parameter $q_0$ describes the behavior of the universe's expansion, quantifying its
acceleration. In our currently accelerating universe, we expect $q_0$ to be negative. The
jerk parameter in turn gives information about inflection points in the expansion history
of the universe. A positive $j_0$ implies that the universe has gone through a change of sign
of the acceleration parameter in the past, meaning that there has been a transition from
deceleration to acceleration.

For our purposes, it is useful to combine the CS among themselves and express~\eqsref{eq:CScoeff}
in terms of each other, yielding
\begin{eqnarray} \label{eq:CSoftime}
	q=-\frac{\dot{\mathcal H}}{\mathcal{H}^2} -1\,, \quad j=\frac{\ddot{\mathcal{H}}}{\mathcal{H}^3}-3q-2\,.
\end{eqnarray}
Moreover, it is worth noting that to convert the time variable $t$ to the redshift $z$, the following
identity
\begin{equation} \label{tempo}
	\frac{dz}{dt}= -\mathcal{H}(z)(1+z) \,
\end{equation}
can be used. One of the most relevant consequences derived from the CS bases on the fact that universe's
energy density can be related to cosmographic parameters without invoking a model \emph{a priori}. This property
has been extensively demonstrated for simple barotropic fluids, in which the pressure is a function of the total
density, i.e. $\mathcal P = \omega \rho $. For a mixture of fluids, in total the net fluid behaves like a fluid
with a general equation of state (EoS) of the form $\omega \equiv \sum_i \mathcal P_i / \sum_i \rho_i$.

We limit our attention to pressureless matter, denoted by the subscript $m$, comprising both baryonic and cold dark
matter, and dark energy. This choice is justified by the fact that present contributions due to neutrinos, photons,
scalar curvature, and so forth are negligible. Hence, we end up with an overall EoS parameter of the
form
\begin{equation} \label{eos}
    \omega=\frac{\mathcal P_{DE}}{\rho_{DE}+\rho_m}\,.
\end{equation}
Using the Friedmann equations with $8\pi G=c=1$,
\begin{subequations}
\begin{align}
  \mathcal H^2 &= \frac{\rho}{3} \,,\\
  \dot{ \mathcal{H}}+\mathcal{H}^2 &= -\frac{1}{6}(3\mathcal P+\rho)\,,
\end{align}
\end{subequations}
and the continuity equation
\begin{equation}
    \frac{d\rho}{dt} + 3\mathcal H \left( \mathcal {P} + \rho \right) =0 \,,
\end{equation}
we obtain the pressure of the universe in terms of $q$ as
\begin{equation}
    \mathcal P = \mathcal H^2 \left( 2 q -1 \right) \,,
\end{equation}
and thus the corresponding expressions for $\omega$ and its first derivative $\omega^{\prime}\equiv \frac{d\omega}{dz} $
in terms of the CS read
\begin{subequations}
\begin{align}
    \omega &= \frac{2q-1}{3}\label{wuz1} \,,\\
    \omega^{\prime}&= \frac{2}{3} \frac{\left( j - q - 2q^2 \right)}{1+z} \label{wuz2}\,.
\end{align}
\end{subequations}
These two quantities give important information about current universe's expansion and recent changes in expansion
behavior. Equations~\eqref{wuz1} and~\eqref{wuz2} show that constraining the current values of $q$ and $j$ corresponds
to fixing limits on the current thermodynamic state of the universe. In other words, possible changes of
cosmographic parameters correspond to changes of the EoS of the universe and its derivatives.
Since we aspire to determine the current values of the CS without the need of assuming an EoS of any given model,
the results of our fits will provide direct constraints on the current EoS parameter $\omega$ and its derivative
$\omega^{\prime}$ by employing~\eqsref{wuz1} and \eqref{wuz2}. \\

Finding numerical values for the CS can be achieved by fitting appropriate data, e.g. the apparent luminosity
of type Ia supernovae as a function of the redshift. It is possible to express the distance modulus $\mu_D$, given
by the difference of the apparent ($\mu_{\rm{app}}$) and absolute luminosity ($\mu_{\rm{abs}}$) of an object, in terms of the luminosity distance $d_L$
as
\begin{equation}
  \mu_D = \mu_{\rm{app}} - \mu_{\rm{abs}} = 25 + \frac{5}{\ln 10} \ln \left( \frac{d_L}{\mathrm{1\,Mpc}} \right) \,.
\end{equation}
and in turn the luminosity distance as a function of the scale factor as
\begin{equation}
	d_L = \mathcal{R}_0 (1+z) = \mathcal{R}_0 \, \frac{1}{a(t)}\,,
\end{equation}
where we used the identity
\begin{equation} \label{eq:redshift}
	a\equiv\frac{1}{1+z}\,.
\end{equation}
Here $\mathcal{R}_0$ is the distance that a photon travels from a light source at $r=\mathcal{R}_0$ to our position
at $r=0$, defined as
\begin{equation}
	\mathcal{R}_0 = \int_{t}^{t_0}{\frac{d\xi}{a(\xi)}}\,.
\end{equation}
We can calculate this quantity by using the power series expansion for the inverse of the scale factor and
integrating each term in the sum separately. \\
By inserting the expansion of $a(t)$ and $\mathcal{R}_0$ into the luminosity distance, we then obtain a Taylor series
expansion of $d_L$. Substituting the time variable $t$ by the redshift $z$ according to~\eqeqref{tempo} it is possible
to obtain the luminosity distance $d_L$, and the distance modulus $\mu_D$ in terms of a Taylor expansion with respect
to the redshift $z$~\cite{2005Viss,2007Catt,2012Avil}. Thus, we obtain for the luminosity distance $d_L$
\begin{eqnarray} \label{eq:dLTaylor}
  d_L &=& d_\mathcal H \, z \, \bigg[ 1 + z \, \left(\frac{1}{2} - \frac{q_0}{2} \right) \\
    &~&~~~~~~~ + z^2 \, \left(-\frac{1}{6} -\frac{j_0}{6} + \frac{q_0}{6} + \frac{q_0^2}{2} \right)
      +...\bigg] \nonumber \,,
\end{eqnarray}
where $d_\mathcal H = 1/\mathcal H_0$ is the Hubble distance. After straightforward calculations, we can obtain the
distance modulus in the form
\begin{eqnarray} \label{eq:muTaylor}
  \mu_D &=& 25 + \frac{5}{\ln 10} \, \bigg[ \ln\left( \frac{d_\mathcal H}{\mathrm{1\,Mpc}} \right) \\
    &~&~~~~~~~ + \ln z +  \zeta_1 \, z + \zeta_2 \, z^2  +... \bigg] \nonumber \,,
\end{eqnarray}
where $\zeta_1$ and $\zeta_2$ are as yet undetermined coefficients. We have to expand the logarithm of the luminosity
distance for small $z$, which leads to the following results for the coefficients: 
\begin{subequations} \label{eq:mudL}
\begin{align}
  \zeta_1 &= \frac{1}{2} - \frac{q_0}{2} \,, \\
  \zeta_2 &= \frac{5 q_0}{12} + \frac{3 q_0^2}{8} - \frac{j_0}{6} -\frac{7}{24}  \,.
\end{align}
\end{subequations}
Equations~\eqref{eq:dLTaylor} and~\eqref{eq:muTaylor} are commonly used in cosmography to fit supernovae data in order to obtain
numerical values for the CS. In the next sections, we will address some problematic issues connected to this formalism,
and in this context introduce the concept of Pad\'e approximations.

\section{The problems with cosmography}
\label{sec:problems}

As previously described, cosmography considers Taylor expansions of relevant observables, which are then constrained
by directly fitting cosmological data. Its methodology permits us to assume that the cosmographic fitting procedure is
\emph{model-independent} from any particular cosmological model, because in any of the expansions used we do not rely
on model-dependent assumptions \emph{a priori}. However, the introduced formalism of determining cosmological bounds
on the CS entails some other difficulties, which have to be addressed. In the following subsections, we describe in
detail each of these problems, which must be alleviated through the use of either theoretical or statistical techniques.

\subsection{Truncated approximations of the Taylor series}

Taylor expansions are approximations to an exact expression, and coincide with the original
function for the limit of infinite terms in the expansion. As it is impossible to consider
an infinite number of terms in numerical analyses, the series has to be truncated at some finite
order. This introduces errors into the analysis, since the formulae
used for fitting only represent approximations to the true expressions.
This problem can be moderated by including higher orders of the series, but this
comes at the expense of introducing more fitting parameters and considerably
complicating the corresponding statistical analysis. Every extension of parameter space
implies a broadening of the posterior distributions for each parameter, and thus in
principle it is desirable to keep the number of fitting parameters as low as possible.

\subsection{Convergence at higher redshifts}

A second issue is related to the range of convergence of Taylor series used for cosmographic
expansions. By definition, the Taylor series we constructed converges only for small $z$. Hence,
it may happen that for higher redshifts the series diverges, being unable to correctly represent the
distance modulus or the luminosity distance for the whole set of cosmic data. SNeIa data reach
up to a redshift of about $z\simeq 1.414$, which lies outside the expected convergence range of
the series; and the inclusion of data from other astrophysical sources can extend the fitting
regime even to redshifts higher than that. Thus, the attempt to improve the statistics of the analysis
by including higher redshift data actually jeopardizes the original aim and leads to a less
accurate result due to the lack of convergence of the fitting functions used. This problem is known
in the literature as \emph{convergence problem}~\cite{25000}.
To avoid divergences, some modifications of the formalism have been proposed, like e.g. rephrasing the
series in terms of a new variable, which compresses the data to a region in which the convergence of
the series is still guaranteed. One possibility for such a new variable, i.e. an alternative definition
of redshift, has been suggested in earlier work~\cite{2005Viss} as $y = z/(1+z)$, which exhibits improved
convergence properties as compared to the conventional redshift $z$. This idea has been extended to further
notions of redshift by~\cite{2012Avil}. The new redshift variable $y$ shows better convergence
behavior in particular in the past redshift regime; for $z \in [0,\infty)$, the new redshift is bound in
$y \in [0,1]$. Less fortunately, in the future regime, it fails to converge, as with $z \in [-1,0]$ we obtain
$y \in (-\infty,0]$. Independently of the exact choice of a new redshift variable, this seems to be a common problem,
i.e. by requiring that possible re-parametrizations must result in a smooth function, the new variable
$z_{new}$, given by $z_{new}=\mathcal{Z}(z)$, with $\mathcal{Z}(z)$ a generic function of the redshift $z$ converging
to $\mathcal Z (z)\rightarrow1$ as $z\rightarrow\infty$, generally shows divergences at $z\rightarrow-1$. Thus, although
the past behavior is of crucial importance, usually the future regime is unfortunately not well constrained and there
the convergence problem remains. We aim at building up an alternative method to alleviate the convergence
problem in both past and future regimes by bringing Pad\'e expansions into play. This idea will be further elaborated
in the following sections.

\section{Principles of Pad\'e expansion}
\label{sec:Pade}

For a function $f(x)$, the Pad\'e approximant of order $(m,n)$ is given by
\begin{equation} \label{facy}
  P_{mn}(x) = \frac{a_0 + a_1 x + a_2 x^2 + ... + a_m x^m}{1+b_1 x + b_2 x^2
    + ... + b_n x^n} \,.
\end{equation}
In principle, it is possible to express the luminosity distance $d_L$ not in form of a
conventional Taylor series $d_L = d_\mathcal H \, z \, \left( 1 + \alpha z + \beta z^2 +... \right)$,
where $\alpha, \beta,...$ are dependent on the parameters of the CS, but in the shape of a Pad\'e
approximant expression.
As well as in the case of a Taylor expansion, the coefficients $a_i,b_i$ depend on the parameters
of the CS. The correspondence between the two functional forms and their sets of parameters, i.e.
expressing the coefficients $a_i,b_i$ by their equivalents $\zeta_1,\zeta_2$, can be obtained
from the requirement that for $z=0$ the two different parametrizations and their derivatives be
equal. In order to achieve a balanced correspondence between Taylor and Pad\'e form, we should
match the number of coefficients, and so we choose $a_0=0$ and look for a Pad\'e approximant of
the form $(1,2)$,
\begin{equation} \label{tropos1}
  d_{\mathrm{L,Pad\acute{e}}} = \frac{a_1 z}{1+b_1 z + b_2 z^2} \,.
\end{equation}
The Pad\'e form of the luminosity distance then reads
\begin{equation} \label{tropos2}
  d_{\mathrm{L,Pad\acute{e}}} = \frac{12 d_\mathcal H z}{12+6(q_0-1)z+(5 + 2 j_0 - 8 q_0 - 3 q_0^2)z^2} \,,
\end{equation}
where
\begin{subequations} \label{eq:aibi}
\begin{align}
a_1 &= d_\mathcal H\,,\\
b_1 &= \frac{1}{2}(q_0-1)\,,\\
b_2 &= \frac{1}{12}(5 + 2 j_0 - 8 q_0 - 3 q_0^2) \,.
\end{align}
\end{subequations}
The thus formulated $d_{\mathrm{L,Pad\acute{e}}}$ cannot diverge, since no real poles occur in the observational
range for $z$ and for any $q_0,j_0$. The Pad\'e-expanded parametrization of the magnitude $\mu_D$ can then be easily
calculated by using~\eqsref{eq:muTaylor} and \eqref{eq:mudL} and results in a rather involved expression due to the
logarithmic form, reported in Appendix~\ref{app:muD}.

While on first sight this may not reveal any improvements with respect to the Taylor expression, some
considerations show that the Pad\'e form, besides some possible drawbacks, does have its advantages.
Within the conventional Taylor treatment, the quantity to be expanded
originally is the scale factor $a(t)$. This expansion defines the cosmographic series to begin with.
Since the luminosity distance is a function of $a(t)$, it can be directly given as a function of the CS. In
the Pad\'e treatment, we have to continue using the Taylor expansion of $a(t)$, since we want to keep
the CS as our parameter set. Expanding $d_L$ into a Pad\'e approximant, and linking the coefficients
$a_i,b_i$ to the CS, thus means to approximate an already approximated expression anew.

To avoid this, it would be necessary to go back to the expansion of the scale factor itself and start with Pad\'e
approximants already at that stage. However, since the CS is defined from a Taylor expansion, and as such offers a
very direct and intuitive interpretation of the dynamics of the universe's expansion, retaining the set of cosmographic
parameters as defined from the conventional Taylor approach is preferable. Following the outlined procedure, the
statistical analyses will show that the re-expression of the Taylor series in terms of Pad\'e approximants does not
significantly propagate systematic errors, and that the use of Pad\'e expansions can be justified retrospectively.

Moreover, in another concern the Pad\'e approximant method has its undeniable advantages, and that is the question
of convergence. Pad\'e approximants are known to have
a much larger radius of convergence, while a Taylor series fails to converge for $z>1$ or even earlier. Since in
usual cosmographic analyses supernova data from a range of redshifts $z\in [0,1.414]$  is used, and the extension of
data sets to information from higher redshift sources is desirable, significant problems with the Taylor formalism
are expected for the high redshift regime of data. With a Pad\'e approximant, it would be no problem to not
only use the full range of supernova data, but to even expand the data sets to include new high redshift sources.
In this context, we regard it as highly useful to consider a cosmographic analysis in the framework
of a Pad\'e expansion.

\subsection{The convergence radius}

In this subsection, we will explicitly demonstrate why the PA represents a better alternative to standard
Taylor expansions. To quantify the advantages of Pad\'e over Taylor formalism, a viable tool is to evaluate
their convergence radii $\mathcal R$. Let us consider the case of a Pad\'e expansion of the
luminosity distance in analogy to~\eqsref{tropos1} and \eqref{tropos2}, but to order $(1,1)$, as a simple
example for evaluating the convergence radius. The choice of a Pad\'e approximant of order $(1,1)$ corresponds
to a second order Taylor series, for which we will calculate the convergence radius as well. Forms of the
luminosity distance with higher orders of expansions would refine the result for the convergence radius,
but not significantly change it. Restating the luminosity distance in the form
\begin{equation} \label{gtshazz}
	d_{L,\rm{Pad\acute{e}}}= d_\mathcal H \frac{\mathcal Az}{\mathcal B+\mathcal C z}\,,
\end{equation}
with the identifications
\begin{subequations}\label{gtshazz2}
\begin{align}
  \mathcal A &= 1\,, \\
  \mathcal B &= 1\,,\\
  \mathcal C &= \frac{1}{2}(q_0-1)\,,
\end{align}
\end{subequations}
we can, by demanding the positivity of the Pad\'e expansion of $d_L$, derive the following condition on
the acceleration parameter:
\begin{equation} \label{eq:q0cond}
	q_0 > -1 \,,
\end{equation}
for the choice $z=1$. This result naturally predicts an accelerated universe in a rather stringent form, but is only
a consequence of restricting the luminosity distance~\eqeqref{gtshazz} in its Pad\'e form to positive values, regardless
of the correct cosmological model. For the sake of clearness, the value presented in~\eqeqref{eq:q0cond} only gives a
lower limit for the acceleration parameter. The Pad\'e expansions thus theoretically exclude $q_0=-1$, which corresponds
to a pure de Sitter universe.

By recalling the definition of the geometrical series for a generic variable $x<1$,
\begin{equation}
	\sum_{n=0}^{\infty} x^n \equiv \frac{1}{1-x} \,,
\end{equation}
and using a reformulation of the luminosity distance obtained after cumbersome algebra,
\begin{equation} \label{eq:dLnew}
  d_{L,\rm{Pad\acute{e}}}=\frac{2}{q_0-1}\Big[1-\frac{2}{2+(q_0-1)z}\Big]\,,
\end{equation}
we can rewrite~\eqeqref{eq:dLnew} in terms of a geometric series as
\begin{equation} \label{eq:dLnew}
  d_{L,\rm{Pad\acute{e}}} = \frac{2}{q_0-1} \left[1 - \sum_{n=0}^{\infty} \frac{z^n}{\left(\frac{2}{1-q_0}\right)^n} \right] \,,
\end{equation}
which converges if the argument of the series is smaller than unity, i.e. if
\begin{equation}
	z < \frac{2}{1-q_0}\,,
\end{equation}
implying that the convergence radius of the Pad\'e expansion is
\begin{equation}
  \mathcal R_{\rm{Pad\acute{e}}} = \frac{2}{1-q_0} \,,
\end{equation}
which for typical values of $q_0 \sim [-0.7,-0.4]$ results in values of $\mathcal R_{\rm{Pad\acute{e}}} \sim [1.18,1.42]$.
Since the regime of convergence of usual cosmographic Taylor expansions has an upper limit of unity, i.e. the highest
convergence radius is $\mathcal R_{\rm{Taylor}} = 1$, we hope to obtain a higher value for $\mathcal R_{\rm{Pad\acute{e}}}$,
which would show that the use of Pad\'e approximants extends the convergence region of the analysis and thus significantly
reduces the problem of convergence plaguing standard cosmography. However, the typical convergence radius of Taylor series
can also be less than the unity. In fact, using the Taylor expansion of $d_L$ up to second order as given by~\eqeqref{eq:dLTaylor},
it can be approximately given as
\begin{equation}
  \mathcal R_{\rm{Taylor}} \simeq \frac{1-q_0}{2} =\mathcal R_{\rm{Pad\acute{e}}}\,,
\end{equation}
which typically lies in the range of $\mathcal R_{\rm{Taylor}} \sim [0.70,0.85]$. \\
In order to obtain numerical values for the expressions of the convergence radii of Pad\'e and Taylor expansions, we should
consult the fitting results for $q_0$. We show the values for the convergence radii in~\tabref{tab:postprocessing}, calculated
from the later obtained results for the CS. We find that, in all cases, the convergence radius of Pad\'e is greater than the
convergence radius of Taylor series, suggesting that the Pad\'e convergence always extends to a larger regime than the one
predicted by standard Taylor formalism.

\section{Fitting results}
\label{sec:results}

In this section, we describe the statistical procedure for fitting the cosmological data through the use of the PAs.
All fits have been carried out for the luminosity $\mu_D$, adopting the union 2.1 compilation~\cite{Suzuki:2011hu} of
distance moduli of supernovae as a function of the redshift $z$, combined with diverse priors on one or more of the
fitting parameters.

\subsection{Supernova data and statistics}

Type Ia supernovae represent well-consolidated rulers for fitting cosmological data, since they are considered
standard candles and have become primary distance indicators. The most recent supernova survey is based on the
union 2.1 compilation of the supernova cosmology project~\cite{Suzuki:2011hu} and spans a wide range of supernovae
in $z \in [0.015,1.414]$. The list of supernovae includes 580 measurements of the distance modulus, extending previous
compilations, e.g. the union 2~\cite{SNeIa-4a} and union 1~\cite{Kowalski:2008ez} sets.

In order to fix cosmographic constraints on the fitting parameters, we use a Monte Carlo analysis employing Markov
chains. The adopted Metropolis algorithm~\cite{metr} enables us to reduce the dependence of the analysis on the initial
statistical distribution by modifying the statistics of the proposal distribution during the run of the Monte Carlo
simulation. In particular, the statistical distribution in one step then depends on the previously used one, leading to
normally distributed numerical results~\cite{metr3}.

In the analyses, one maximizes the following likelihood function
\begin{equation} \label{funzionea}
	{\mathcal L} \propto \exp(-\chi^2/2)\,,
\end{equation}
where $\chi^2$ is the chi-square function, defined as
\begin{equation}
	\chi^{2} =\sum_{k=0}^{580}\frac{(\mu_{D,k}^{\mathrm{th}}-\mu_{D,k}^{\mathrm{obs}})^{2}}
	{\sigma_{k}^{2}}\,,
\end{equation}
with $\mu_{D,k}^{\mathrm{th}}$ the theoretical value of $\mu_D$, predicted by the fitting function used, and
$\mu_{D,k}^{\mathrm{obs}}$ the observed luminosities given by data catalogs. The corresponding $1\sigma$ error
to each supernova is denoted by $\sigma_k$ and is reported by the supernova survey as well.
The investigated ranges of parameters adopted for the analyses are
\begin{subequations} \label{ranges12}
\begin{align}
  h &\in [0.4,0.9]\,,\\
  q_0 &\in [-1.5,0]\,,\\
  j_0 &\in [-2,2]\,,
\end{align}
\end{subequations}
where $h$ is defined via $\mathcal H_0 = 100\, h \, \mathrm{km/(s\,Mpc)}$. These ranges agree with theoretical
predictions \cite{2008Wein,2005Viss,2007Catt,2012Avil,25000} and are in agreement with the Planck results~\cite{Planck}.
Different analyses were performed without and then with the assumptions of priors imposed on the fitting parameters. For nearly
all of the fits we used the complete range of data with redshifts $z\in [0,1.414]$, while one fit was carried out for a
restricted sample of supernova data with $z\in [0,0.36]$. Further, one fit was performed using Taylor parametrization to
show the differences between the two approaches.

\subsection{Cosmological priors}

Cosmological priors are used in order to simplify or concretize the numerical analysis.
In fact, it is possible that using a single fitting function depending on many parameters
is not enough to separately or sufficiently constrain all parameters due to occurring
degeneracies among them. By fixing viable priors on parameters, the total phase
space is reduced, and complementing different priors with each other may lead to new insights
and compelling results for the fitting parameters. For those reasons, we carried out our numerical
analyses with and without adopting numerical priors. For $\mathcal H_0$, two different priors were
imposed; one being obtained from the best fit for $\mathcal H_0$ extracted from the most recent
observations by the Planck collaboration~\cite{Planck}, i.e. $\mathcal H_0 = 67.11\,\mathrm{km/(s\,Mpc)}$.
The second was obtained by fitting the union 2.1 supernova data in the range $z<0.36$, expanding
the luminosity distance into a power series to first order, where the Taylor and the Pad\'e
approach coincide, i.e.,
\begin{equation}
  d_L \simeq \frac{1}{\mathcal H_0} \, z \,.
\end{equation}
This procedure results in a value of $\mathcal H_0 = 69.96_{-1.16}^{+1.12}\,\mathrm{km/(s\,Mpc)}$,  which was used as a prior
in one of the fits. \\
Furthermore, an additional prior on $q_0$ may be imposed assuming that the $\Lambda$CDM model is the
limiting case of a more general paradigm, using a constant DE term combined with the Planck results.
In a universe containing baryonic and dark matter with the density $\Omega_m$ and a cosmological
constant with the density $\Omega_{\Lambda} = 1-\Omega_m$, the Hubble parameter evolves as
\begin{equation}
  \mathcal H = \mathcal H_0\sqrt{\Omega_m(1+z)^3+1-\Omega_m} \,,
\end{equation}
leading to an acceleration parameter of
\begin{equation}
  q_0 =  -1 + \frac{3}{2}\Omega_m \,.
\end{equation}
Hence, by using the result from the Planck mission for the matter density, $\Omega_m = 0.3175$, the prior on the
acceleration parameter turns out to be $q_0 = -0.5132$. This numerical value for $q_0$ is used for one of the fits
performed, which we can then compare to the results obtained without fixing any parameter \emph{a priori}, with the
ones where a prior on $\mathcal H_0$ alone is used, and the one where $\mathcal H_0$ and $q_0$ are both fixed.

As mentioned before, we performed several fits, distinguished by numbers in~\tabref{tab:dL}. Without priors, we carried
out analyses using the Taylor approach (1), the Pad\'e parametrization (2), and the Pad\'e parametrization
using the short redshift range $z\in [0,0.36]$ (3). Further, we presumed priors from Planck's results on $\mathcal H_0$
only (4), on $q_0$ only (5) and on both $\mathcal H_0$ and $q_0$ (6), as well as a prior on $\mathcal H_0$ from the first-order
fit of the luminosity distance (7), for a total of seven different fits. For each fit, the respective $p$-values have
been reported in~\tabref{tab:dL}, representing the probabilities that the result obtained by a single fit is observed,
supposing the null hypothesis to hold. Since it is a qualitative measure for the likelihood of a certain outcome of a fit,
it is expected to be as close as possible to unity~\cite{verde,verde2}.

\begin{figure}
	\begin{center}
	\includegraphics[width=0.45\textwidth]{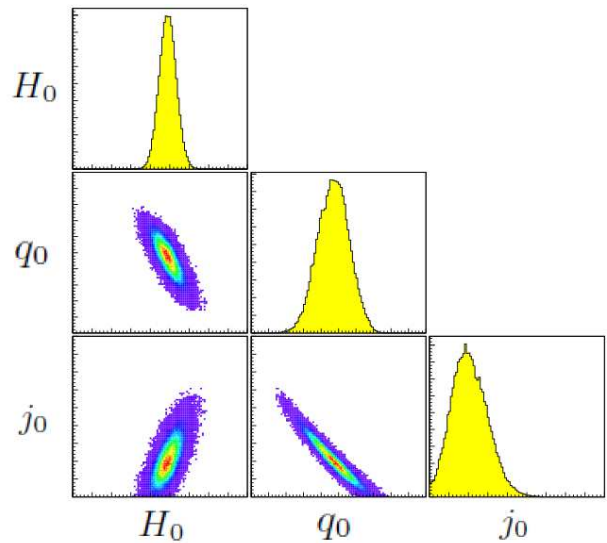}
	{\small \caption{(color online) Contour plots and posterior distributions
	for $\mathcal H_0$, $q_0$ and $j_0$, for a fit with Taylor parametrization
	and without any priors imposed. }
	\label{fig:Taylor_all}}
	\end{center}
\end{figure}

\begin{figure}[ht]
	\begin{center}
	\includegraphics[width=0.45\textwidth]{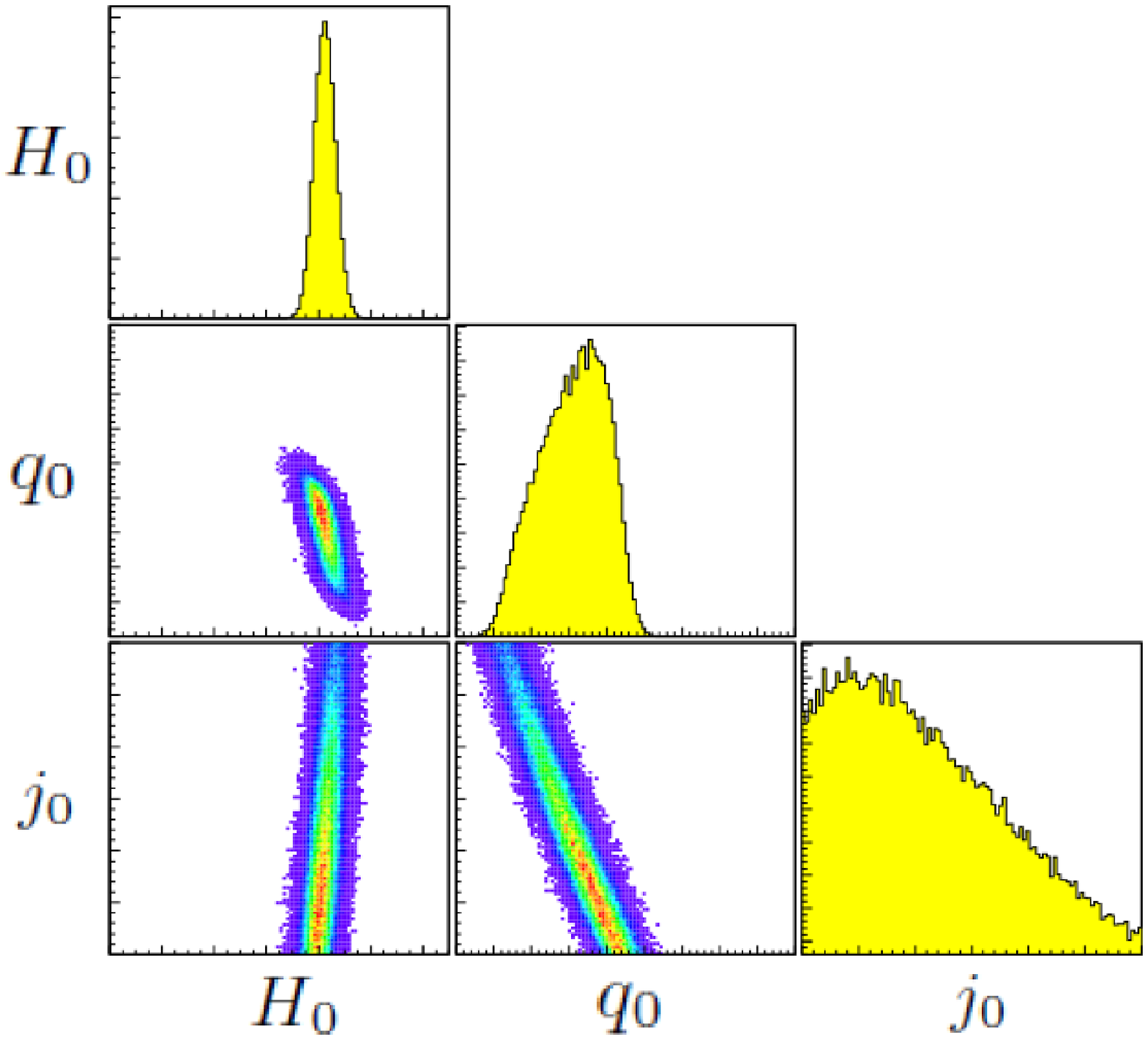}
	{\small \caption{(color online) Contour plots and posterior distributions
	for $\mathcal H_0$, $q_0$ and $j_0$, for a fit with Pad\'e parametrization
	and without any priors imposed. }
	\label{fig:Pade_all}}
	\end{center}
\end{figure}

\begin{figure}[ht]
	\begin{center}
	\includegraphics[width=0.45\textwidth]{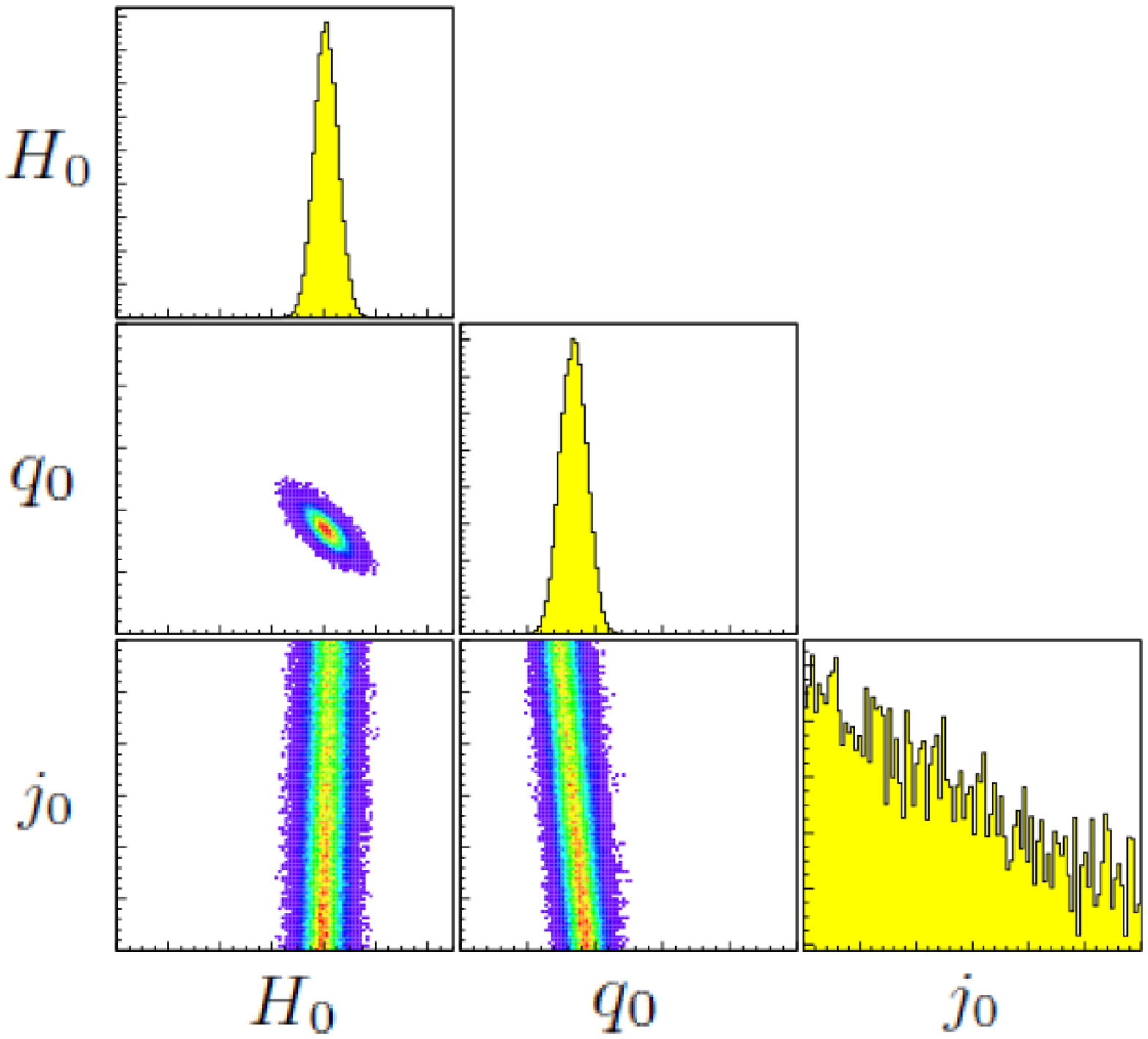}
	{\small \caption{(color online) Contour plots and posterior distributions
	for $\mathcal H_0$, $q_0$ and $j_0$, for a fit with Pad\'e parametrization
	and without any priors imposed, for the short redshift range.}
	\label{fig:Pade_all_short}}
	\end{center}
\end{figure}

\begin{figure}[ht]
	\begin{center}
	\includegraphics[width=0.33\textwidth]{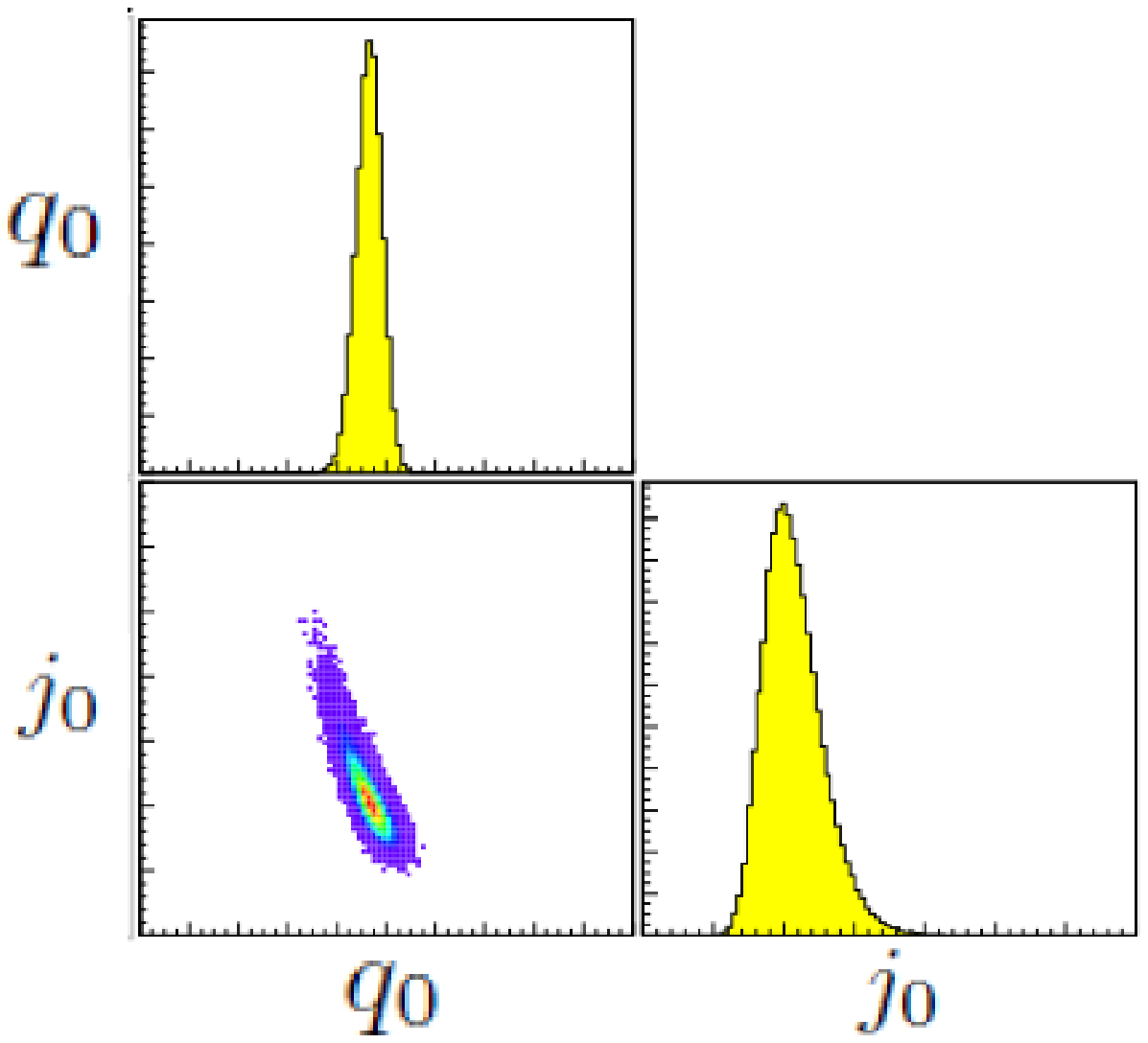}
	{\small \caption{(color online) Contour plots and posterior distributions
	for $\mathcal H_0$, $q_0$ and $j_0$, for a fit with Pad\'e parametrization
	and with a prior on $\mathcal H_0$ from the Planck results. }
	\label{fig:Pade_H0_Planck}}
	\end{center}
\end{figure}

\begin{figure}[ht]
	\begin{center}
	\includegraphics[width=0.35\textwidth]{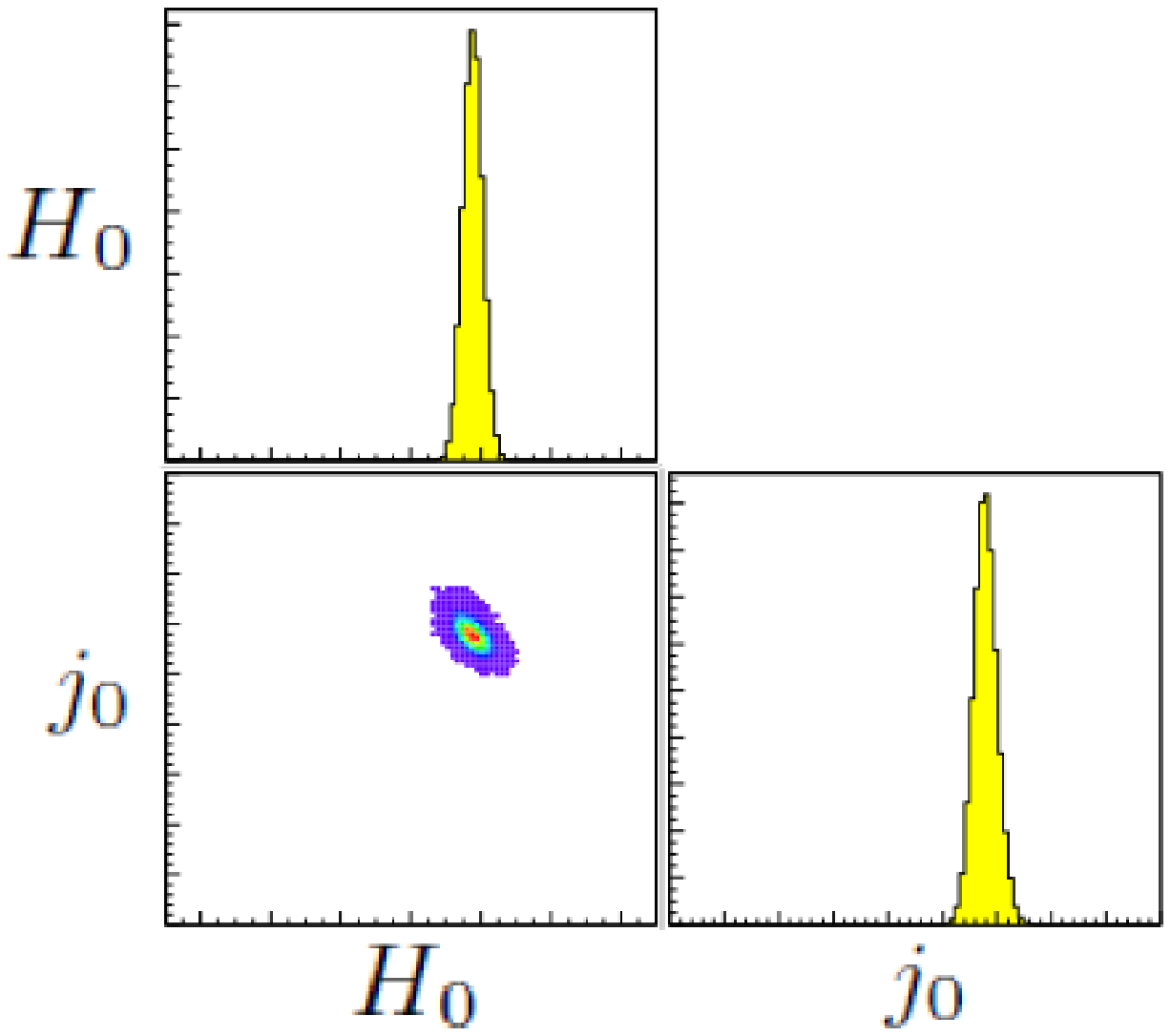}
	{\small \caption{(color online) Contour plots and posterior distributions
	for $\mathcal H_0$, $q_0$ and $j_0$, for a fit with Pad\'e parametrization
	and with a prior on $q_0$ from the Planck results.}
	\label{fig:Pade_q0_Planck}}
	\end{center}
\end{figure}

\begin{figure}[ht]
	\begin{center}
	\includegraphics[width=0.33\textwidth]{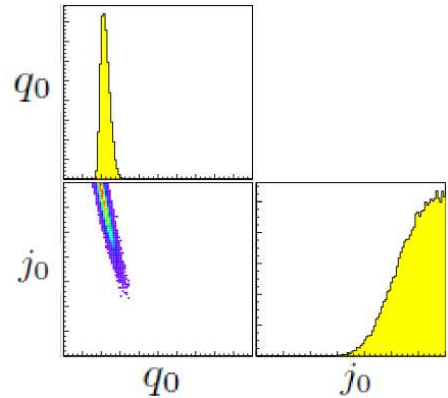}
	{\small \caption{(color online) Contour plots and posterior distributions
	for $\mathcal H_0$, $q_0$ and $j_0$, for a fit with Pad\'e parametrization
	and with a prior on $\mathcal H_0$ from the first order fit of the luminosity
	distance.}
	\label{fig:Pade_H0_Pade}}
	\end{center}
\end{figure}

All the numerical results for the CS have been reported in~\tabref{tab:dL}, with the presumption of priors indicated in
each column, and the corresponding contour and distribution plots in Figures~\ref{fig:Taylor_all}-\ref{fig:Pade_H0_Pade}.

The results of the fits are as different as the priors that have been involved in the analysis.
The implications from the obtained $p$-values also permit us to conclude that some hypotheses are
disfavored strongly in comparison to others. In particular, from the fits (1) and (2), where no priors
have been used, we can draw a comparison of the Taylor and the Pad\'e treatments. Using Taylor
expansions leads to a slightly lower Hubble parameter than derived from the Pad\'e treatment, as well
as a less negative acceleration
parameter and a much smaller jerk parameter. Overall, the results from the Taylor treatment
resemble much more the $\Lambda$CDM predictions, but predict a significantly higher value of
the Hubble parameter than $\Lambda$CDM. The $p$-values of (1) and (2) are nearly equal, albeit
$p$ for fit (2) is slightly higher. The outcomes of the results of fit (3), which was carried
out for the restricted sample of redshifts, support the results from the Pad\'e approach in
fit (2), but predict a slightly smaller Hubble parameter than (2), whereas the acceleration
and jerk parameter are rather close to the values of (2). The $p$-value of (3) is however
significantly larger, which can be accounted for by the smaller sample of redshifts used, leading
to higher accuracy of the result. The similarity of (2) and (3) indicates however the validity
of the Pad\'e approach regardless of the regime of redshifts used in the analysis. \\

From the results of fits (4) and (5), it seems that the imposition of priors leads to rather
disadvantageous results. The predicted values of $q_0$ and $j_0$ in fit (4) do not make much
sense, since $q_0$ is barely negative, which indicates an only slowly expanding universe, and
$j_0$ is negative, which is in contradiction with
the assumption of changes in the direction of expansion in the cosmological history.
Correspondingly, the $p$-value of this fit seems to indicate a very low likelihood for these
results. From fit (5), only imposing the Planck prior on $q_0$, the predicted Hubble parameter
is again larger than expected from the Planck data analysis, and $j_0$ again negative, with a
surprisingly large $p$. Fixing both $\mathcal H_0$ and $q_0$ finally, the obtained jerk parameter
is very large and positive, but at the disadvantage of a very low $p$.
However, generally $p$-values below a certain significance level are only taken as an indication
for the rejection of the null hypothesis, i.e., the assumption of no connection between the input
and the outcome for the fitting parameter, therefore the particularly low $p$ seems to indicate
a deeper connection of the three parameters $\mathcal H_0$, $q_0$ and $j_0$, and not a particularly bad
fitting likelihood. \\
Ultimately, the last fit using the prior on the Hubble parameter obtained from the first-order
fit of the luminosity distance seems to be the most successful of the analysis. The Hubble
parameter is higher than predicted by $\Lambda$CDM, but in close accordance with the results
from fits (2) and (3). Furthermore, the results for acceleration and jerk parameter are both
reasonable simultaneously, $q_0$ being slightly less negative than the values predicted from
(2) and (3), but more negative than the Planck prior; whereas the jerk parameter is positive
and nearly identical to $j_0=1$ as predicted by the $\Lambda$CDM model. The $p$-value of fit
(6) is the highest of all, and identical to $p$ from fit (2). \\
In summary, the results seem to indicate that the current Hubble parameter is higher than the
value claimed in the cosmological analyses by the Planck collaboration~\cite{Planck}, which can
be inferred from the results of the (statistically favorable) fits (2), (3) and (7), as well as
from the poor results in fits (4) and (5). The acceleration parameter is comparable to the one
predicted by the $\Lambda$CDM model, while the jerk parameter is positive, although tendentially
larger than the $\Lambda$CDM value $j_0=1$~\cite{orly}.

\begin{table*}
\caption{{\small Table of best fits and their likelihoods (1$\sigma$) for the parameters
	  $\mathcal H_0$, $q_0$ and $j_0$.}}

\begin{tabular}{c|c|c|c|c|c|c|c}
\hline \hline

\hline \hline

\hline

fit 			& fit (1) & fit (2) & fit (3) & fit (4) & fit (5) & fit (6) & fit (7) \\ [0.8ex]
\hline
$p$-value		& {\small $0.690$ }	& {\small $0.694$ } &   {\small $0.811$ } &   {\small $0.242$ }
 			& {\small $0.689$ } 	& {\small $0.019$ } &   {\small $0.694$ } \\[0.8ex]
\hline
{\small$\mathcal H_0$}		& ~{\small $69.90$}{\tiny${}_{-0.433}^{+0.438}$}~	& ~{\small $70.25$}{\tiny${}_{-0.403}^{+0.410}$}~
			& ~{\small $70.090$}{\tiny${}_{-0.450}^{+0.460}$}~	& ~{\small $67.11$}~
			& ~{\small $69.77$}{\tiny${}_{-0.290}^{+0.288}$}~	& ~{\small $67.11$}~
			& ~{\small $69.96$}{\tiny${}_{-1.16}^{+1.12}$}~ \\[0.8ex]

{\small$q_0$}		& ~{\small $-0.528$}{\tiny${}_{-0.088}^{+0.092}$}~	& ~{\small $-0.683$}{\tiny${}_{-0.105}^{+0.084}$}~
			& ~{\small $-0.658$}{\tiny${}_{-0.098}^{+0.098}$}~	& ~{\small $-0.069$}{\tiny${}_{-0.055}^{+0.051}$}~
			& ~{\small $-0.513$}~ 					& ~{\small $-0.513$}~
			& ~{\small $-0.561$}{\tiny${}_{-0.042}^{+0.055}$}~ \\[0.8ex]

{\small$j_0$}		& ~{\small $0.506$}{\tiny${}_{-0.428}^{+0.489}$}~	& ~{\small $2.044$}{\tiny${}_{-0.705}^{+1.002}$}~
			& ~{\small $2.412$}{\tiny${}_{-0.978}^{+1.065}$}~	& ~{\small $-0.955$}{\tiny${}_{-0.175}^{+0.228}$}~
			& ~{\small $-0.785$}{\tiny${}_{-0.208}^{+0.220}$}~ 	& ~{\small $2.227$}{\tiny${}_{-0.237}^{+0.245}$}~
			& ~{\small $0.999$}{\tiny${}_{-0.468}^{+0.346}$}~ \\[0.8ex]

\hline \hline

\end{tabular}

{\tiny Note. $\mathcal H_0$ is given in Km/(s\,Mpc).

}

\label{tab:dL}
\end{table*}

\subsection{Implications for the EoS parameter}

The EoS parameter $\omega$ and its first derivative with respect to the redshift $\omega^{\prime}$
can be directly inferred from the fitting results for the CS via expressions~\eqref{wuz1} and~\eqref{wuz2},
which were introduced in~\secref{sec:cosmography}. At the present time, we can evaluate the two quantities using
the different values of $q_0$ and $j_0$ obtained in the numerical analyses by
\begin{eqnarray}
	\omega_0 &=& \frac{2q_0-1}{3} \,, \\
	\omega^{\prime}_0 &=& \frac{2}{3} \left(j_0 - q_0 - 2 q_0^2 \right) \,.\nonumber
\end{eqnarray}
The results can be found in~\tabref{tab:postprocessing}. Disregarding fit (4), which has
proven to lead to dubious results, also regarding the EoS, the values of the EoS parameter vary in the
expected range of $\omega_0 \in [-0.789,-0.675]$. The results from fits (5) and (6) with the Planck
priors on $q_0$ are the least negative, whereas those from fits (2) and (3) have larger absolute
values. It seems that in general cosmography predicts a slightly more negative EoS than the one predicted
by the $\Lambda$CDM model,
\begin{equation} \label{w0a}
  \omega_0=-1+\Omega_m\,,
\end{equation}
which, with an overall matter density of $\Omega_m = 0.314 \pm 0.02$ is constrained in the interval
$\omega_0\in[-0.688,-0.684]$. The variation of the EoS parameter given by its first derivative is clearly
positive in all cases except fits (4) and (5), which confirms that the universe is in a state of transition
between different equations of state, evolving towards a more negative EoS parameter in the future~\cite{ratra}.
In this regard, the $\Lambda$CDM model predicts
\begin{equation} \label{w0all}
  \omega^{\prime}_0=3(1 - \Omega_m) \Omega_m\,,
\end{equation}
which lies in the range $\omega^{\prime}_0\in[0.644,0.648]$. \\
The results inferred for $\omega^{\prime}_0$ show that models with negative constant pressure seem to be favored
in describing the acceleration of the universe. However, degeneracies between models occur, since besides the
concordance $\Lambda$CDM there exist others, as e.g. the one proposed in~\cite{io}, where the dark energy density
evolves in time, but has constant negative pressure and a vanishing speed of sound. From the above fitting results,
it is not possible to distinguish between those two models. Summarizing, the statistical fitting results are all
compatible with the $\Lambda$CDM model, however, they do not exclude other models as e.g. ones with a varying DE
term and vanishing speed of sound~\cite{io,ss,ss2,dp39,2012Avil}.

\begin{table*}
\caption{{\small Table of convergence radii for the Pad\'e expansion of the luminosity distance,
as well as the EoS parameter $\omega_0$ and its first derivative $\omega'_0$ at
current time, for diverse results of $q_0$ and $j_0$ obtained.}}

\begin{tabular}{c|c|c|c|c|c|c}
\hline \hline

\hline \hline

\hline \quad fit \quad &
\quad\quad {\small $q_0$} \quad\quad & \quad\quad {\small $j_0$} \quad\quad &  \quad\quad $\mathcal R_{\rm{Pad\acute{e}}}$ \quad\quad &
	\quad\quad $\mathcal R_{\rm{Taylor}}$ \quad\quad & \quad\quad $\omega_0$ \quad\quad & \quad\quad $\omega^{\prime}_{0}$ \quad\quad \\ [0.8ex]
\hline
    ~~fit (1)~~ &
    ~{\small $-0.528$}{\tiny${}_{-0.088}^{+0.092}$}~ & ~{\small $0.506$}{\tiny${}_{-0.428}^{+0.489}$}~
    & ~{\small $1.309$ }{\tiny${}_{-0.075}^{+0.079}$}~ & ~{\small $0.764$}{\tiny${}_{-0.044}^{+0.046}$}~
    & ~{\small $-0.685$}{\tiny${}_{-0.059}^{+0.061}$}~ & ~{\small $0.317$}{\tiny${}_{-0.293}^{+0.333}$}~ \\ [0.8ex]
\hline
    ~~fit (2)~~ &
    ~{\small $-0.683$}{\tiny${}_{-0.105}^{+0.084}$}~ & ~{\small $2.044$}{\tiny${}_{-0.705}^{+1.002}$}~
    & ~{\small $1.188$}{\tiny${}_{-0.074}^{+0.059}$}~  & ~{\small $0.842$}{\tiny${}_{-0.052}^{+0.042}$}~
    & ~{\small $-0.789$}{\tiny${}_{-0.07}^{+0.056}$}~ & ~{\small $1.196$}{\tiny${}_{-0.485}^{+0.675}$}~ \\ [0.8ex]
\hline
    ~~fit (3)~~ &
    ~{\small $-0.658$}{\tiny${}_{-0.098}^{+0.098}$}~  & ~{\small $2.412$}{\tiny${}_{-0.978}^{+1.065}$}~
    & ~{\small $1.206$}{\tiny${}_{-0.071}^{+0.071}$}~ & ~{\small $0.829$}{\tiny${}_{-0.049}^{+0.049}$}~
    & ~{\small $-0.772$}{\tiny${}_{-0.065}^{+0.065}$}~ & ~{\small $1.469$}{\tiny${}_{-0.661}^{+0.718}$}~ \\ [0.8ex]
\hline
    ~~fit (4)~~ &
    ~{\small $-0.069$}{\tiny${}_{-0.055}^{+0.051}$}~  & ~{\small $-0.955$}{\tiny${}_{-0.175}^{+0.228}$}~
    & ~{\small $1.870$}{\tiny${}_{-0.096}^{+0.089}$}~  & {\small $0.535$}{\tiny${}_{-0.027}^{+0.025}$}~
    & ~{\small $-0.380$}{\tiny${}_{-0.036}^{+0.034}$}~ & ~{\small $-0.597$}{\tiny${}_{-0.12}^{+0.154}$}~ \\ [0.8ex]
\hline
    ~~fit (5)~~ &
    ~{\small $-0.513$}~  & ~{\small $-0.785$}{\tiny${}_{-0.208}^{+0.220}$}~
    & ~{\small $1.322$}~ & ~{\small $0.757$}~
    & ~{\small $-0.675$}~  & ~{\small $-0.532$}{\tiny${}_{-0.138}^{+0.147}$}~ \\ [0.8ex]
\hline
    ~~fit (6)~~ &
    ~{\small $-0.513$}~   & ~{\small $2.227$}{\tiny${}_{-0.237}^{+0.245}$}~
    & ~{\small $1.322$}~  & ~{\small $0.757$}
    & ~{\small $-0.675$}~ & ~{\small $1.476$}{\tiny${}_{-0.158}^{+0.164}$}~ \\ [0.8ex]
\hline
     ~~fit (7)~~ &
    ~{\small $-0.561$}{\tiny${}_{-0.042}^{+0.055}$}~ & ~{\small $0.999$}{\tiny${}_{-0.468}^{+0.346}$}~
    & ~{\small $1.281$}{\tiny${}_{-0.034}^{+0.045}$}~ & ~{\small $0.780$}{\tiny${}_{-0.021}^{+0.027}$}~
    & ~{\small $-0.707$}{\tiny${}_{-0.028}^{+0.037}$}~ & ~{\small $0.620$}{\tiny${}_{-0.314}^{+0.235}$}~ \\ [0.8ex]

\hline \hline

\end{tabular}

\label{tab:postprocessing}
\end{table*}

\section{Conclusions and outlook}
\label{sec:concl}

In this paper we used a model-independent procedure based on cosmography to fix constraints
on cosmological parameters in order to investigate observational data. In particular, in order to fit the
CS, i.e. a set of observables characterizing the kinematics of the universe, we introduced
a new technique to parameterize the fitting functions in form of Pad\'e expansions, extending
conventional treatments with Taylor series used in cosmography. \\
The Pad\'e approach turns out
to be advantageous in the reduction of systematic weaknesses of the Taylor treatment, alleviating the
convergence problem associated to the truncation and the range of validity of a Taylor series being
limited to the regime $z<1$. We demonstrated the derivation of Pad\'e expansions from the well-known
Taylor expressions, and provided the parametrization of the distance modulus $\mu_D$, as well as the luminosity distance
$d_L$ in terms of the CS. Solely from the expression for the luminosity distance in Pad\'e form, it is possible
to infer the condition $q_0 > -1$ on the acceleration parameter, which is not a consequence of fitting
cosmological data, but a mere theoretical prediction originating from the definition of the luminosity
distance. Further, we defined the convergence radius of the Pad\'e expansion in terms of the CS,
to be calculated from the fitting results and compared to the convergence regime of Taylor series. \\

Fits were carried out using the union 2.1 compilation of the supernova cosmology project, and with fitting functions
in Taylor and Pad\'e parametrizations. We adopted a Monte Carlo analysis with Markov chains implementing a Metropolis
algorithm. We used different or no priors on the fitting parameters $\mathcal H_0$, $q_0$
and $j_0$, and obtained the best fit values including the $1\sigma$ errors. In general, the results seem
to indicate a larger Hubble parameter and a slightly more negative acceleration parameter than the ones
found by the Planck mission~\cite{Planck}. Further, a clearly positive jerk parameter has been found, implying
a transition in the expansion dynamics of the universe at a finite past redshift.

The imposition of priors from the Planck mission leads to rather conflictive results and statistically unfavorable
fitting behaviors, whereas the prior on $\mathcal H_0$ from a first-order expansion of the luminosity distance produced
one of the two statistically best fits, the other one being a fit without any priors assumed. Comparing the outcome of
fits with Taylor and Pad\'e approaches, the Taylor treatment yields results closer to the Planck predictions, but with
a slightly lower $p$-value than the corresponding Pad\'e approach. This can be directly interpreted as an improvement
of the convergence problem.

From the results for $\mathcal H_0$, $q_0$ and $j_0$ we further calculated the current values of the EoS parameter $\omega_0$
and its first derivative $\omega^{\prime}_0$ at the present time. The results allow for the possibility of the EoS of the universe
being in a transition between a matter-dominated state and a de Sitter expansion, and indicate a decreasing EoS parameter in the
future. Fairly good agreement with the $\Lambda$CDM predictions has been found, although none of the results permit us to conclude
with certainty that the DE density is constant in time. In fact, our results do not exclude \emph{a priori} a varying DE term with
vanishing speed of sound, since our numerical bounds comply with the theoretical ranges predicted by such a model~\cite{io,ss,ss2}. \\

Summarizing, we introduced Pad\'e expansions as a new technique to perform cosmographic fits, yielding improved results with respect
to the standard Taylor treatment, in terms of convergence radii and data analyses. Our approach represents a viable alternative to
standard methods and shows good statistical fitting behaviors, yielding bounds compatible with Planck's first results. These investigations
will be object of future efforts and more precise analyses, by expanding series to higher orders and refining the Pad\'e expansions
by working with even more accurate observational datasets, which would provide further insights on the form of both the EoS of DE and its
evolution in time, reconstructing its functional behavior as the universe expands.

\acknowledgements

O.L. is grateful to M. Scinta for his support during the course of this research, and wishes to express his gratitude
to prof. S. Capozziello for discussions. C.G. is supported by the Erasmus Mundus Joint Doctorate Program by
Grant Number 2010-1816 from the EACEA of the European Commission.



\begin{widetext}

\appendix
\section{Pad\'e expansion of the distance modulus}
\label{app:muD}
The distance modulus formulated in terms of a Pad\'e approximant as a function of the redshift $z$ reads
\begin{equation}
  \mu_{\mathrm{D,Pad\acute{e}}} = \frac{5}{\ln 10} \left[ \ln z + \frac{\mathcal D}{\mathcal E} \right] \,,
\end{equation}
where
\begin{subequations}
\begin{align}
  \mathcal D &= a_0 + a_1 z \,,\\
  \mathcal E &= b_0 + b_1 z + b_2 z^2 \,,
\end{align}
\end{subequations}
and
\begin{subequations}
\begin{align}
  a_0 =& ~-24 \bigg[ -6 -35 \ln 10 - 20 \,j_0\, \ln 10 + q_0^2 \left( -6 + 45 \ln 10 \right) +
      2\,q_0 \left( 6+ 25 \ln 10 \right) \\
    & + \left( 9\,q_0^2 + 10\,q_0 - 4\,j_0 -7 \right) \ln \left(\frac{d_{\mathcal H}}{1 \rm{Mpc}}\right) \bigg]
      \left[ 5\ln 10 + \ln \left(\frac{d_{\mathcal H}}{1 \rm{Mpc}}\right) \right]\,,\nonumber\\
  a_1 =&~ 24 \left( q_0 -1\right) \bigg[ -3 -35 \ln 10 - 20 \,j_0 \ln 10 + q_0^2 \left( -3 + 45 \ln 10 \right)
      + q_0 \left( 6+50 \ln 10 \right) \\
    & + \left( 9\,q_0^2 + 10\,q_0 - 4\,j_0 -7 \right) \ln \left(\frac{d_{\mathcal H}}{1 \rm{Mpc}}\right) \bigg] \,,\nonumber\\[1mm]
  b_0 =&~ 24 \left(4 \,j_0-9 \,q_0^2-10 \,q_0+7\right) \ln \left(\frac{d_{\mathcal H}}{1 \rm{Mpc}}\right)
      + 480\, j_0 \ln 10+144\, q_0^2-1080 \,q_0^2 \ln 10 \\[2mm]
    & -288 \,q_0-1200 \,q_0 \ln 10 + 144+840 \ln 10 \,,\nonumber\\[2mm]
  b_1 =&~ -12 \left(4 \,j_0+17\right) q_0+48 \,j_0+108 \,q_0^3+12 \,q_0^2+84 \,,\\[2mm]
  b_2 =&~ 16 \,j_0^2-2 \left(36 \,j_0+13\right) q_0^2-20 \left(4\, j_0+7\right) q_0+56\, j_0+81\, q_0^4+180 \,q_0^3+49 \,.
\end{align}
\end{subequations}

\end{widetext}

\end{document}